\documentclass[pra,preprint,amssymb,showpacs]{revtex4}
\usepackage{amsmath}
\usepackage{dcolumn}
\usepackage{bm}
\usepackage{graphicx}
\def\be{\begin{equation}}
\def\ee{\end{equation}}
\def\bea{\begin{eqnarray}}
\def\eea{\end{eqnarray}}
\def\source#1#2#3#4{{\it #1}~{\bf #2}, #3 (#4)}
\def\Eq#1{Eq.~\ref{#1}}

\def\Ref#1{Ref. \cite{#1}}

\def\CA{{\cal A}}
\def\CB{{\cal B}}
\def\CC{{\cal C}}
\def\CD{{\cal D}}
\def\CM{{\cal M}}
\def\CN{{\cal N}}
\def\CR{{\cal R}}

\def\al{\alpha}
\def\om{\omega}
\def\irt{\frac{1}{\sqrt{3}}}

\def\GHZ#1{|\Psi_#1 \rangle}
\def\ket#1{| #1 \rangle}
\def\bra#1{\langle #1 |}

\def\BX{{\bf X}}

\begin{document}
\title{Many-qutrit Mermin inequalities with three measurement settings}
\author{Jay Lawrence} 
\affiliation{Department of Physics and Astronomy, Dartmouth
          College, Hanover, NH 03755, USA}
\affiliation{The James Franck Institute, University of Chicago, 
          Chicago, IL 60637}
\date{revised \today}
\bigskip
\begin{abstract}
Mermin inequalities are derived for systems of three-state particles (qutrits)
employing three local measurement settings.  These establish perfect 
correlations which violate local realistic bounds more strongly than those 
previously reported with two bases.  The quantum eigenvalue of this Mermin 
operator grows as the dimension of the Hilbert space, $3^N$, rather than 
$2^N$, as obtained with two bases.  
The number of distinct GHZ contradictions also increases as $3^N$.

\end{abstract}
\pacs{03.67-a, 03.65.Ta, 03.65.Ud}
\maketitle
\section{Introduction}
This work is motivated by recent experimental breakthroughs on the entanglement of
three nonbinary particles  \cite{Malik.16, Erhard.18}.  In particular, a three-qutrit GHZ 
state has recently been produced and documented for the first time \cite{Erhard.18}.  
The qutrits are realized with photon orbital angular momentum, and years of progress 
leading to the present breakthroughs are reviewed in \Ref{twisted}.  Here we build 
upon a recent theoretical calculation for qutrits \cite{JL17} by extending the number 
of local measurement settings from two to three, equal to the number of outcomes.  
This extension enhances the violations of local realism qualitatively. 
  
Such violations are expressed by Greemberger-Horne-Zeilinger (GHZ) paradoxes 
\cite{GHZ}, or by Bell inequalities \cite{Bell64,CHSH}.  In this work we choose a special
form of the latter, called Mermin inequalities, because they reveal individual GHZ 
contradictions implicitly, as pointed out in Mermin's original work \cite{Mermin.90,EPR}.  
The Mermin {\it operator} is a sum of $N$-particle tensor products, each of which has 
the same given state as an eigenstate, with the same eigenvalue.  Hidden variables 
can duplicate this eigenvalue for a subset of these tensor products, but are thereby
constrained to predict different values for others.  This gives rise to specific GHZ 
contradictions as well as to violations of the Mermin inequality itself, which places 
an upper bound on the hidden variable value of the Mermin operator.  A Mermin 
inequality has been defined formally by Cabello et. al. \cite{Cabello.08}, as a Bell 
inequality (I) whose ``Bell operator is a sum of stabilizing operators'' that represent 
the perfect correlations in their simultaneous eigenstate, and (II) which maximizes 
the violations for that state.  Here as in \cite{JL17}, we have taken this definition to 
include {\it concurrent} operators \cite{LLK}, which do not commute but share a 
common eigenstate - these were essential to the discovery of GHZ contradictions 
\cite{Ryu13,JL14} and Mermin inequalities \cite{JL17} for qutrits. This work shows 
that concurrent operators built from three local measurement settings achieve an 
eigenvalue growth rate $\sim3^N$, the appropriate analog to Mermin's $\sim 2^N$ 
for qubits, achieved with commuting stabilizers.

A brief review of general developments since GHZ \cite{GHZ} and Mermin 
\cite{Mermin.90} was given in \Ref{JL17} and won't be repeated here.  But we should 
emphasize, for the sake of comparison, that Bell inequalities have been developed 
which serve as criteria for {\it irreducible} $N$-particle entanglement - so that violations 
for a given state rule out its factorization into any subsets of fewer than all $N$ particles 
\cite{Svetlichny.87,Collins.02}.  A relevant example is a recent treatment of qutrit 
systems \cite{Alsina.16} in which, as here, three measurement settings were used.  
The Bell operators are more complex and require separate analyses for different 
numbers of particles.   Qutrit results are presented for the cases of $N = 2 - 6$.  As 
expected from earlier examples, the violations of this type of Bell inequality grow 
more slowly with $N$ than the violations of Mermin inequalities.  




In the next section we describe a set of related GHZ states and the rotationally 
covariant observables that comprise the Mermin operator, and we compare its 
quantum and classical values.  The quantum values will be obvious, but the 
classical maxima require proof, and this is spelled out in Section III.  In the 
concluding section, we discuss the physical significance of our results, including
a count of GHZ contradictions, and we comment on higher dimensional systems.

\section{The Mermin operator - quantum and classical values}

To construct the Mermin operator, first consider the nine related $N$-qutrit 
GHZ states,
\be
   \GHZ{k} =  \irt \big( \ket{00...0} + \al^k \ket{11...1} + \al^{2k} \ket{22...2} \big),
   \hskip1.0truecm          (k = 0,1,...,8),
\label{GHZstates}
\ee
where $\al = \exp(2\pi i/9)$, so that each $\ket{\Psi_k}$ is generated from 
$\ket{\Psi_{0}}$ by a rotation through $2 \pi k/9$, as shown in Fig. 1a.  It is a defining 
symmetry of GHZ states \cite{JL14} that such rotations may be distributed arbitrarily 
among qutrits (about their respective $\hat{z}$ axes), in increments that add up to 
the net rotation angle.

\begin{figure}
\includegraphics[scale=.65]{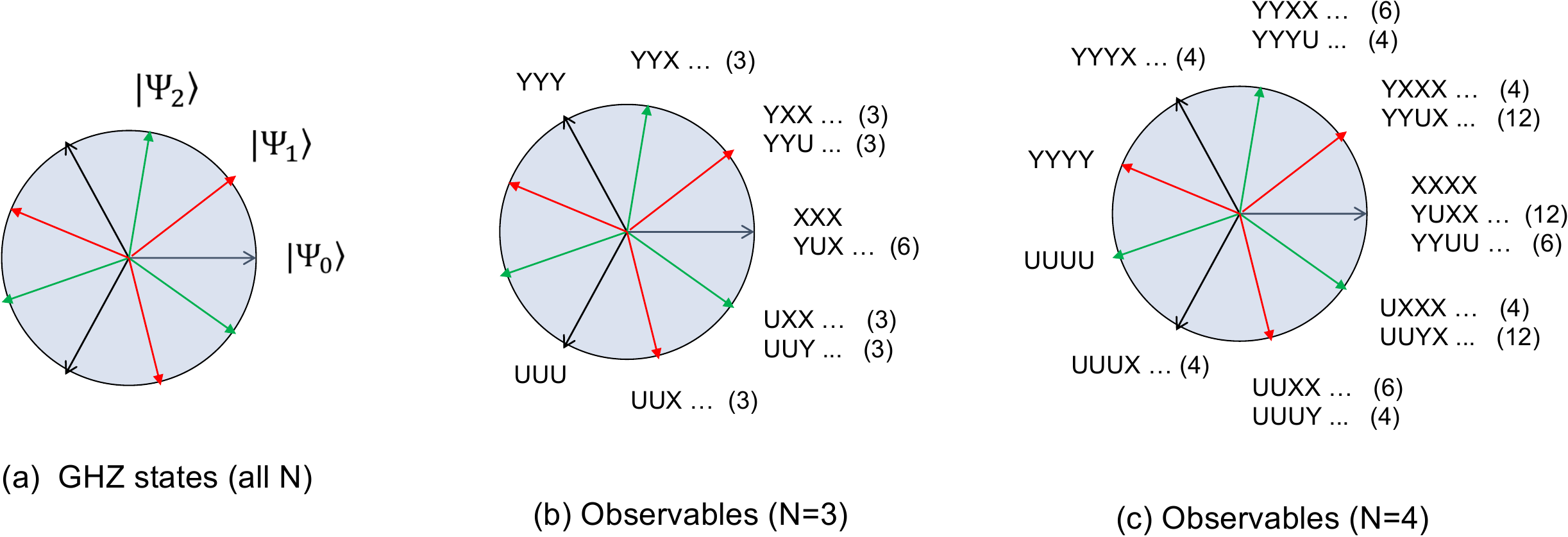}
\caption{\label{fig1} (a) GHZ states (\Eq{GHZstates}), and (b) tensor product 
observables for $N=3$ and (c) $N=4$.   Parentheses denote the number of 
permutation-related tensor products.   Black arrows define the subset whose joint 
eigenstate is $\GHZ{0}$.   Red and green arrows relate similarly to $\GHZ{1}$ and 
$\GHZ{2}$.}
\end{figure}

Next consider an observable of which $\GHZ{0}$ is an eigenstate with eigenvalue
unity, 
\be
   \BX \equiv X^{\otimes N} = X_1...X_N.
\label{BigX}
\ee
Its factors are the standard qutrit Pauli matrices,  
$X_i = \sum_{n=0}^2 \ket{n+1}_i \bra{n}_i$, acting on the i{\it th} qutrit.   Rotations of 
these factors through the basic angles ($\pm 2\pi/9$) generate other qutrit matrices 
which we call, respectively (dropping the index $i$),
\bea
    &  Y \equiv Z^{1/3} X Z^{-1/3} =  \sum_{n=_0}^2 \ket{n+1} \al^{(1 - 3\delta_{n,2})}
         \bra{n},    \\
    &  V \equiv Z^{-1/3} X Z^{1/3} =  \sum_{n=_0}^2 \ket{n+1} \al^{(3\delta_{n,2} -1)}       
         \bra{n};
\label{Ymatrix}
\eea
where $Z = \sum_{n=0}^2 \ket{n} \om^n \bra{n}$ is the usual diagonal Pauli matrix, which 
rotates qutrits through $2 \pi/3$ about $\hat{z}_i$.   Given $X$, $Y$, and $V$ factors for
each qutrit, we construct the $3^N$ tensor product operators in which each factor can 
be $X$ or $Y$ or $V$.   Every such tensor product is generated from $\BX$ by some 
combination of one-qutrit rotations (through $0$ or $\pm 2 \pi/9$, respectively), and 
Figs. 1(b and c) show the net rotations for $N=3$ and 4.  Because of rotational 
covariance, all operators at point $k$ share the state $\ket{\Psi_k}$ as an eigenstate 
with eigenvalue unity.  

Now, operators have a periodicity property \cite{JL14} - a rotational Bloch theorem:   
If any factor ($X$, $Y$, or $V$) is rotated through $2 \pi/3$, it is simply multiplied by 
$\om$; that is, $ZXZ^{-1} = \om X$, and similarly for $Y$ and $V$.  This means that 
$\om XXX$ (like $YYY$) appears at the point 3 (rotation angle $2 \pi/3$), and it follows 
that every operator at point 3 has $\ket{\Psi_0}$ as an eigenstate {\it with eigenvalue} 
$\om$.  Similarly, every operator at point 6 has $\ket{\Psi_0}$ as an eigenstate with 
eigenvalue $\om^2$.  Therefore, we may define the Mermin operator for the state 
$\ket{\Psi_0}$ as
\bea
  &    \CM_0 =  \hbox{(sum of operators at}~k=0)   \nonumber    \\
  +   & \om^2 \hbox{(sum of operators at}~k=3) + 
                \om \hbox{(sum of operators at}~k=6),
\label{M0general}
\eea
which includes all operators identified by black arrows in Fig. 1, and in which every 
term contributes $+1$ to the eigenvalue, so that
\be
   \CM_0 \ket{\Psi_0} = 3^{N-1} \ket{\Psi_0}.
\label{eigenvalue}
\ee
One could define different Mermin operators, $\CM_1$ and $\CM_2$, corresponding to
$\ket{\Psi_1}$ and $\ket{\Psi_2}$ and identified by red and green arrows, respectively.
These have identical eigenvalues because each accounts for one-third of all operators 
appearing on the plot.  We focus on $\CM_0$ because its higher symmetry simplifies 
the analysis.
 
Let us briefly compare the above quantum result with the classical, or hidden variable
result.  The assumption embodying local realism, or noncontextuality, is that every local
factor, $X_i$, $Y_i$, $V_i$, takes a definite value [eg, $v(X_i) = 1$, $\om$, or $\om^2$], 
and it must take the same value wherever it appears.   A given choice produces a 
classical value of the Mermin operator, $v(\CM_0)$.  Our goal is to find its maximum 
absolute value, $\big| v(\CM_0) \big|$, over all such choices, which we call $\CM_C$.   
We shall simply state here (and prove in the following section) that the maximum is realized 
when all local factors take the same value (eg., unity).  Then, \Eq{M0general} reduces to 
\bea
   &  \CM_{C} = (\hbox{number of operators at}~k = 0)  
    \nonumber    \\
      +  & \om^2 \hbox{(number of operators at}~k=3) + 
                  \om  \hbox{(number of operators at}~k=6) 
   \nonumber    \\
   = &  (\hbox{number of operators at}~k = 0) - (\hbox{number of operators at}~k=3~or~6),
\label{maxHV}
\eea   
where $\om^2 + \om = -1$ and the equality of numbers at $k=3$ and 6 was used.   The 
resulting values are compared in Table I with the quantum eigenvalue, $\CM_Q = 3^{N-1}$.    

\begin{table}[tbp]
\caption{Quantum eigenvalues, $\CM_Q$, classical upper bounds, $\CM_{C}$, and 
their ratio, $\CR$, for $N$ qutrits.  We also list the number of distinct GHZ contradictions 
which may be constructed from terms appearing in the Mermin operator.}
\begin{equation*}
\begin{tabular}{|c|ccc|c|}
\hline
  \  $N$  \  &  \    $\CM_Q$  \  &  \  $\CM_{C}$  \  &  \  $\CR$ \ & \ $\CN_{GHZ}$ \\   \hline
   \   3   \  &   \   9     \ &  \  6  \  &   \   1.5  \  & \   2   \\
   \   4   \  &   \   27   \ &  \   15   \  &  \   1.8 \   & \   8  \\
   \   5   \  &   \   81   \  &  \   36   \   &  \   2.25  \  & \  30     \\
   \  6    \  &   \   243  \ &  \  90  \   &  \   2.70  \  & \ 102     \\
   \   7   \  &   \   729  \ &  \  225  \   & \   3.24  \  & \ 336   \\
 \hline
\end{tabular}
\end{equation*}
\end{table}

Each violation of Mermin's inequality (ratio $\CR \equiv \CM_Q/\CM_{C}  >1$) 
reflects many specific GHZ contradictions (or paradoxes), which may be extracted from 
Tables 1b,c and their larger-$N$ generalizations.  Reference \cite{JL14} shows that HV 
assignments respecting the (common) eigenvalue of a subset of operators at the $k=0$ 
point predict the wrong value for any operator at point 3 or point 6 \cite{JL14b}.  The 
number of distinct GHZ contradictions is equal to the number of wrong predictions, so that
\be
   \CN_{GHZ} = \hbox{(number of operators at}~k=3~and~6) =  
   \frac{2}{3} \big(\CM_Q - \CM_C \big),
\label{GHZC}
\ee
as illustrated in Table I.   The second equality follows from Eqs. \ref{M0general} - 
\ref{maxHV}.  A detailed discussion of these and similar GHZ contradictions will be 
given elsewhere.


\section{Proof of classical maxima}

We now prove that \Eq{maxHV} indeed provides the maximum hidden variable value 
of $\CM_0$, and we derive a closed-form expression.   We begin by showing that the 
Mermin operator is given by the identity,
\bea
    &  \CM_0 = \frac{1}{3} \bigg[\bigotimes_{i=1}^N (X_i+\al^2 Y_i+\al^{-2} V_i) 
        + \bigotimes_{i=1}^N (X_i+\om \al^2 Y_i +\om^2 \al^{-2} V_i)    \nonumber   \\
      &  + \bigotimes_{i=1}^N (X_i+\om^2 \al^2 Y_i +\om \al^{-2} V_i) \bigg].
\label{M0identity}
\eea
To verify, note that each product generates a weighted sum of all $3^N$ operators shown
on the circle graph.   Then, using $\om^2 + \om + 1 = 0$, one can show that a given term 
survives in the sum only if the number of $Y$ factors equals the number of $V$ factors mod. 3.  
This locates surviving terms at the black arrows in Fig. 1.  Finally, using $\al^3 = \om$, one can
show that the multiplying factors of these terms are just those given by \Eq{M0general}.  
Thus, \ref{M0identity} is equivalent to \ref{M0general}. 

We shall evaluate the HV value, $v(\CM_0)$, directly from \ref{M0identity}.  $v(\CM_0)$ is 
a function of the values, $v(X_i)$, etc., assigned to each local factor.  But its magnitude,
$|v(\CM_0)|$, depends only on two independent local ratios, which we choose to be
\be
   R_i = v(Y_i)/v(X_i)  \hskip1.0truecm    \hbox{and}      \hskip1.0truecm  
   S_i = v(V_i)/v(X_i).
\label{ratios}
\ee
We then have  
\bea
     &  \big| v(\CM_0;R_i,S_i) \big| = \frac{1}{3}  \bigg| \bigotimes_{i=1}^N (1+\al^2 R_i +\al^{-2} S_i)
       + \bigotimes_{i=1}^N (1 + \om \al^2 R_i +\om^2 \al^{-2} S_i)    \nonumber   \\
     &  + \bigotimes_{i=1}^N (1 + \om^2 \al^2 R_i +\om \al^{-2} S_i) \bigg|   
      \label{M0value}    \\
     & \equiv \frac{1}{3}  \bigg| \bigotimes_{i=1}^N \CB(R_i,S_i) +
     \bigotimes_{i=1}^N \CC(R_i,S_i) + \bigotimes_{i=1}^N \CA(R_i,S_i) \bigg|,
   \label{M0HVvalue}
\eea
where the last line simply assignes names to the individual qutrit factors in the expression above.
These names are chosen because, when evaluated at $R_i = S_i = 1$ (the ``uniform HV" point), 
the sums become
\bea
    &  \CA(1,1) = (1 + \om^2 \al^2 + \om \al^{-2}) =  1 + 2 \cos {2 \pi \over 9}  \equiv A
    \approx 2.532,       \label{Anumber}    \\
   &  \CB(1,1) = (1 + \al^2 + \al^{-2}) = 1 + 2 \cos {4 \pi \over 9} \equiv B
   \approx 1.347,        \label{Bnumber}    \\
   &  \CC(1,1) = (1 + \om \al^2 + \om^2 \al^{-2}) =  1 +2 \cos{8 \pi \over 9} \equiv -C
   \approx - 0.879;      \label{Cnumber}    
\eea
their magnitudes are ordered as $A > B > C > 0$, with $C(1,1) \equiv -C$ being negative.
The three complex contributions to each of $A$, $B$, and $-C$ are shown in Fig. 2.  
\begin{figure}
\includegraphics[scale=.65]{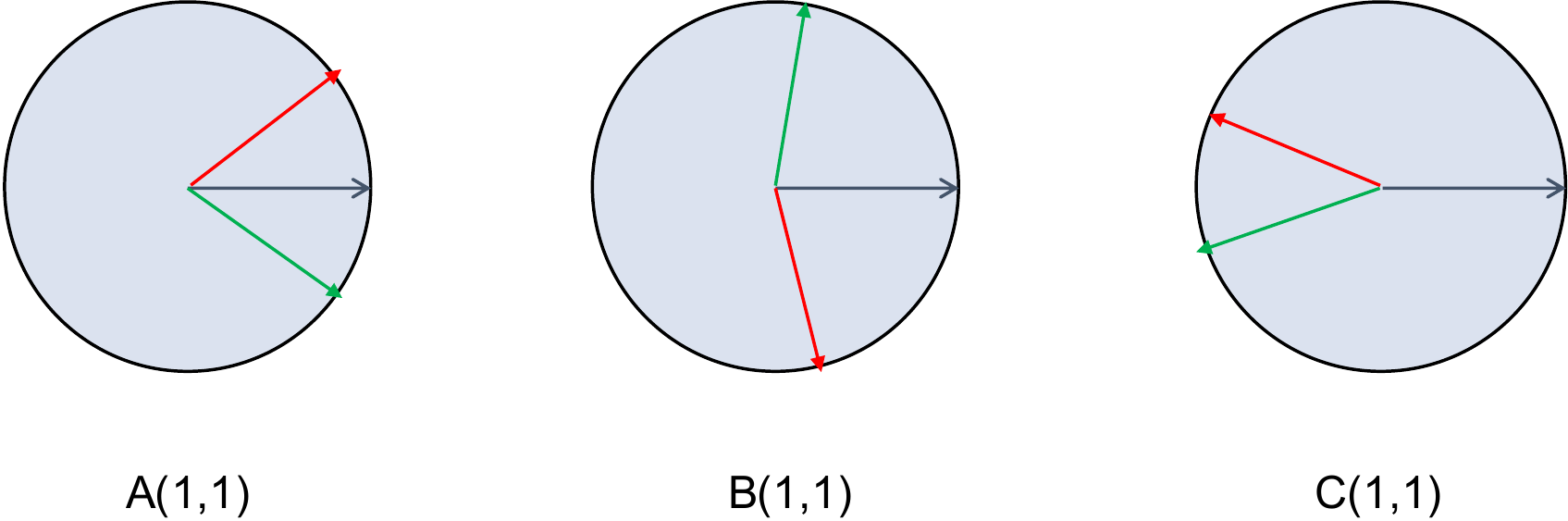}
\caption{\label{fig1} Complex contributions to each of the real numbers $A$, $B$, and $-C$ 
appearing in Eqs. \ref{Anumber} - \ref{Cnumber}, respectively.  Phase angles are 
color-coded as in Fig. 1.}
\end{figure}

The hidden variable magnitude of $\CM_0$ at this ``uniform HV'' point  is given by
\be
  \big| v(\CM_0;1,1) \big|= \frac{1}{3}  \big(A^N + B^N \pm C^N \big),   \hskip1.5truecm  
  \hbox{for}  \hskip 0.2truecm    N~\hbox{even/odd},
\label{HVmax}
\ee
which duplicates the results of \Eq{maxHV} and Table I.  We now show that this 
expression gives the maximum possible value of $|v(\CM_0;R_i,S_i)|$ for all $N \geq 3$.   

\begin{table}[tbp]
\caption{Dependence of single-qutrit factors $\CA(R_i,S_i)$, etc., on hidden variable choices.  
$A$, $B$, and $C$ are real positive numbers (Eqs. \ref{Anumber} - \ref{Cnumber}), and phase 
angles are given in degrees where needed.}
\begin{equation*}
\begin{tabular}{|cc|ccc|}
\hline
  \  $R$  \  &  \  $S$  \  &  $\CA(R,S)$    &   $\CB(R,S)$    &    $\CC(R,S)$  \\   \hline  \hline
   \ \   1   \ \   &   \ \ 1 \ \  &  \ \  $A$   \ \ &  \ \  $B$   \ \ &  \ \  $-C$   \ \         \\  \hline
   \   $\omega$   \  &   \ \  1   \ \  &  $A(40^o)$  &  $B(-80^o)$ & \ $C(-20^o)$     \\
   \ \  1   \ \  &   \  $\omega^2$  \ &  $A(-40^o)$ &   $B(80^o)$ &  $C(20^o)$    \\  \hline
   \   $\omega$   \  &   $\omega^2$  &  \ \  $B$   \ \  &  \ $-C$ \  & \ \ $A$  \ \    \\
      $\omega^2$   &  \ $\omega$ \  &  \   $-C$  \   &  \ \  $A$  \ \  & \ \  $B$  \ \    \\  \hline
      $\omega^2$  &  \ \ 1 \ \   &  $C(20^o)$  &  $A(-40^o)$  &  $B(80^o)$  \\
   \ \  1  \ \    &   $\omega$  &  $C(-20^o)$  &  $A(40^o)$  &  $B(-80^o)$  \\
   \   $\omega$   \  &   \ $\omega$ \  & $B(80^o)$   &  $C(20^o)$  & $A(-40^o)$   \\
      $\omega^2$   &  $\omega^2$  \  & $B(-80^o)$   &  $C(-20^o)$  & $A(40^o)$   \\
\hline
\end{tabular}
\end{equation*}
\end{table}

As background, Table II shows how the hidden variable choices ($R$ and $S$) affect the
individual qutrit factors appearing in \Eq{M0HVvalue}.  To derive Table II, observe that the  
choices $R = \om$ ($\om^2$) correspond, in Fig. 2, to rotations of each {\it green} arrow by 
120$^o$ ($-120^o$), while $S = \om$ ($\om^2$) correspond to rotations of each {\it red} 
arrow by 120$^o$ ($-120^o$).  All entries follow immediately.   Two entries are pure 
rotations in the complex plane.  Two others are pure cyclic permutations.  The remaining 
four are combinations of both.  But note that every choice preserves the total number of 
times ($N$) that each factor ($A$, $B$, $C$) appears in the {\it sum} in \Eq{M0HVvalue}.   
So, under an arbitrary set of HV assignments ($R_i,S_i$), expression \ref{HVmax} will 
be modified by a reshuffling of these factors among the three terms, preserving the total 
number of each factor; and a rotation of each term in the complex plane.  From this alone,
it is clear that expression \ref{HVmax}, for {\it even} $N$, gives the maximum of 
$|v(\CM_0,R_i,S_i)|$ over all hidden variable choices.

The odd $N$ case requires further discussion:  We must show that no HV assignment can 
realign the $C^N$ term without a compensating reduction of $A^N + B^N$.   It is easy to 
show this for any assignment producing a net permutation.  For example, in the case of 
$N=3$, no such trial value $|v(\CM_0,R_i,S_i)|$ can exceed $A^2B + B^2C + C^2A 
\approx 4.06$, which is less than $A^3 + B^3 - C^3 = 6$.   The failure is more dramatic for 
larger $N$, for any set of one-qutrit permutations that does not include all of the particles 
(which simply reproduces the maximum value).

It remains to consider the pure rotations.  Table II shows that there are two possible 
outcomes:   The $A^N$ and $B^N$ terms can either be aligned (zero relative phase), or 
not (relative phases $\pm 120^o$).  The latter case is clearly ruled out.  In the former case, 
the $C^N$ term is always oppositely aligned for odd $N$, so that \Eq{HVmax} indeed 
represents the maximum value, $\big| v(\CM_0;1,1) \big|= \CM_{HVM}$, for all $N$.  

It is clear from \Eq{HVmax} that the large-$N$ asymptote is 
$\CM_{C} \rightarrow \frac{1}{3}(2.532)^N$, so that the quantum to classical ratio 
increases as
\be
    \lim_{N \rightarrow \infty} (\CM_Q/\CM_{C}) \approx (3/2.532)^N \approx 1.185^{N},
\label{limit}
\ee
as compared with $1.064^{N}$ when two measurement settings are used.  Also note that
the divergence of this ratio shows that  \Eq{GHZC} reduces, asymptotically, to
\be
   \lim_{N \rightarrow \infty} (\CN_{GHZ}/ \CM_Q) = 2/3.
\label{GHZC2}
\ee

 
\section{Conclusions}

We have derived many-qutrit Mermin inequalities employing three measurement settings.  
The effect of this extension is shown in Table III, which compares the qutrit cases (two 
{\it vs} three settings), with Mermin's original qubit proof \cite{Mermin.90}.   In the two 
cases where the number of independent measurement settings ($s$) is equal to the 
particle's dimension $d$, the quantum value $\CM_Q$ grows as the dimension $\CD$ 
of the Hilbert space of the system.  It is well known that $\CD-1$ is the number of operators 
which can be diagonalized simultaneously.  It is also the number of operators which can 
share a common eigenstate.   So $\CM_Q$, the number of operators which contribute to 
$\CM_0$, grows as a fixed fraction of the total number that can take sharp values 
simultaneously.   In the qubit case, 
$\CM_Q = \CD/2$, while in the qutrit case, $\CM_Q = \CD/3$.  The same is true, 
asymptotically, for the number of distinct GHZ contradictions, where 
$\CN_{GHZ}  \rightarrow \CD/4$ for qubits \cite{JL05}, while 
$\CN_{GHZ}  \rightarrow 2 \CD/9$ for qutrits (Eq. \ref{GHZC2}).  

\begin{table}[tbp]
\caption{Mermin eigenvalue $\CM_Q$ compared with the dimension $\CD$ of the Hilbert 
space.  $d$ is the particle dimension and $s$ is the number of measurement settings.}

\begin{equation*}
\begin{tabular}{|c|cc|cc|}
\hline
\ source  \ & \ $d$  \  &  \ $s$ \ & \   $\CM_Q$  \  &  \  $\CD$  \\   \hline
\ qubits (\Ref{Mermin.90})  \ & \   2  \ & \  2 \ &  \  $2^{N-1}$  \  &  \  $2^N$    \\
\ qutrits  (\Ref{JL17})   \ & \  3   \  &   \   2   \ &  \ $2^N/3$ \  &  \  $3^N$     \\
\  qutrits (present)  \ & \  3   \  &   \   3   \  &  \   $3^{N-1}$ \  &  \  $3^N$     \\
 \hline
\end{tabular}
\end{equation*}
\end{table}

The symmetry of the construction in Fig. 1 extends to systems of higher odd dimensions, 
provided that $s = d$ independent settings are used.  On this
basis, for $d = 5$, we would expect the quantum value $\CM_Q = 5^{(N-1)} = \CD/5$, 
although the hidden variable maximum is less clear-cut.   Supposing that the optimal 
HV assignment is the uniform one, we could guess the asymptotic classical maximum 
to be $\CM_{C} \approx 4.6898^{N}/5$, so that the quantum violation of this bound 
grows as $\CR \approx 1.066^N$.  

There are daunting practical limitations to preparing entangled systems with larger
$d$ as well as larger $N$.  Regarding fundamental limitations, however, there is no
limit on $N$, while increasing $d$ narrows the quantum-classical gap ($\CR$).  
The narrowing found here suggests that hidden variables can better mimic the 
quantum values as $d$ increases.
\end{document}